\documentclass[12pt,preprint]{aastex}
\def\mearth{M_\oplus}

\shorttitle{Jupiter and Saturn formation}
\shortauthors{Alibert et al.}

\begin{document}

\title{New Jupiter and Saturn formation models meet observations}

\author{Yann Alibert,$^{1}$ Olivier Mousis,$^{1,2}$ Christoph Mordasini,$^{1}$ \& Willy Benz$^{1}$}

\altaffiltext{1}{Physikalisches Institut, University of Bern, Sidlerstrasse 5, CH-3012 Bern, Switzerland, and ${}^2$Observatoire de Besan\c{c}on, CNRS-UMR 6091, BP 1615,
25010
Besan\c{c}on Cedex, France. email: Yann.Alibert@phim.unibe.ch, Olivier.Mousis@obs-besancon.fr, Christoph.Mordasini@phim.unibe.ch, Willy.Benz@phim.unibe.ch.}

\clearpage

\begin{abstract}

The wealth of observational data about Jupiter and Saturn provides
strong constraints to guide our understanding of the formation of
giant planets. The size of the core and the total amount of heavy
elements in the envelope have been derived from internal structure
studies by Saumon \& Guillot (2004). The atmospheric abundance of some volatile elements
has been measured {\it in situ} by the {\it Galileo} probe 
(Mahaffy et al. 2000, Wong et al. 2004) or by remote sensing
(Briggs \& Sackett 1989, Kerola et al. 1997). In this Letter, we show that, by extending
the standard core accretion formation scenario of giant planets by Pollack et al. (1996)
to include migration and protoplanetary disk evolution, it is
possible to account for all of these constraints in a self-consistent
manner.
\end{abstract}


\clearpage

\section{Introduction}
\label{intro}

The standard giant planet formation scenario is the so-called
core-accretion model. In this model, a solid core is formed first by
the accretion of solid planetesimals. As the core grows, it eventually
becomes massive enough to gravitationally bind some nebular gas,
forming a gaseous envelope in hydrostatic equilibrium. The further
increase in core and envelope masses lead to larger and larger radiative
losses which ultimately prevent the existence of an equilibrium envelope.
Runaway gas accretion occurs, rapidly building up a giant planet.
This scenario, which naturally implies giant planets enriched in heavy
elements compared to the Sun, has suffered so far from the problem that
the resulting formation time is comparable to, or longer than, the
lifetime of protoplanetary disks as inferred from observations (Pollack et al. 1996 - hereafter P96, Haisch et al. 2001).
One approach to solve this long-standing problem has been to
revise the opacities used to model the planet's envelope (see Hubickyj et al. 2003)
and/or to allow for a local enhancement in the number of planetesimals
(Klahr \& Bodenheimer 2003). However, Alibert et al. (2004,2005a - hereafter A04 and A05a) have shown recently that extending the
original core accretion scenario to include migration of the growing
planet and protoplanetary disk evolution, results in a formation
speed-up by over an order of magnitude even without local density
enhancements or modified opacities\footnote{The capability of migration to prevent the isolation
of the protoplanetary core was already stated by Ward (1989) and Ward \& Hahn (1995).}. In this Letter,
we show that in addition to solving the formation timescale problem,
these models can also account for the characteristics of the two most
well known giant planets: Jupiter and Saturn.

Within the framework of our model, we find that the uncertainties on
the characteristics of the initial protoplanetary disk are large enough
to allow us to match the observed properties of a single planet (Jupiter
for example) relatively easily. The situation is more complicated with
{\it two} planets forming within the {\it same} protoplanetary disk. For
each satisfactory model matching the observed properties of Jupiter (total
mass, distance to the sun, mass of the core, total mass of heavy elements
and volatiles enrichments), only two parameters are left in order to
account for the same five quantities in Saturn: the initial location of
the embryo and the time offset (that can be equal to 0) between the start
of the formation of both planets. The purpose of this Letter is to show
that, by assuming reasonable properties for the initial protoplanetary
disk, it is possible to construct models of Jupiter and Saturn,
compatible with all the observations detailled in the next two paragraphs.

Using measurements of Jupiter and Saturn (mass, radius, surface temperature,
gravitational moments, etc.) and state-of-the-art structure
modeling, Saumon \& Guillot (2004 - hereafter SG04) have derived important constraints regarding the possible
internal structure of these planets.  From this modeling,
$M_{\rm core}$, the mass of the core of the planet and $M_{\rm Z,enve}$,
the amount of heavy elements in the envelope (assumed to be homogeneously
distributed) can be obtained. In the case of Jupiter, the  maximum total
amount of heavy elements present in the planet  ($M_{\rm core} + M_{\rm Z,enve}$)
is of the order of $\sim 42 M_\oplus$ (Earth masses), whereas the mass of
the core can vary from $0$ to $13 M_\oplus$. This large uncertainty
is essentially due to the undertainties in the equation of state of hydrogen.
In the case of Saturn, $M_{\rm Z,enve}$ ranges from  nearly 0 to 10 $M_\oplus$,
the mass of the core being between  8 and 25 $M_\oplus$.
Note however that
the mass of the solid core might be reduced by up to $\sim 7 M_\oplus$
depending upon the extend of sedimented helium, a process which is required
to explain the present day luminosity of the planet (Fortney \& Hubbard 2003, Guillot 2005).

Abundances of some volatile species in the atmosphere of Jupiter have been measured
using the mass spectrometer on-board the {\it Galileo} probe (see Mahaffy et al. 2000, Wong et al. 2004).
These measurements show that the planet's atmosphere is enriched
in C, N, S, Ar, Kr, and Xe by a factor of $3.7 \pm 0.9$, $3.2 \pm 1.2$, $2.7
\pm 0.6$, $1.8 \pm 0.4$, $2.4 \pm 0.4$, and $2.1 \pm 0.4$ respectively
compared to solar values (Lodders 2003). For Saturn, ground-based observations
(Brigg \& Sackett 1989, Kerola et al. 1997) have shown that C and N are enriched by a factor of $3.2 \pm 0.8$
and $2.4 \pm 0.5$ compared to the solar values. Since the two planets are
almost entirely convective, we assume that these enrichments are representative
of the mean envelope composition.

In Sect. 2 of this Letter, we give a short presentation of our formation models.
In Sect. 3, we apply these models to Jupiter and Saturn, and show an important role
of Jupiter's formation on that of Saturn. In Sect. 4, we calculate
the enrichments in volatile species in the atmosphere of the two planets, and Sect. 5 is devoted to summary
and conclusions.

\section{Formation models}

Our formation models consist in the simulation of the time evolution of
the protoplanetary disk and of the two planetary seed embryos that will
eventually lead to Jupiter and Saturn. We calculate in a consistent way
the structure and evolution of the disk,
the migration of the planets, and their growth in mass
due to accretion of gas and planetesimals.

The evolution of the protoplanetary disk is calculated in the framework of
the $\alpha$ formalism (Shakura \& Sunyaev 1973). 
The initial gas surface density $\Sigma$ inside the protoplanetary disk
(which extends from 0.25 AU to 30 AU) is given by $\Sigma \propto r^{-3/2}$.
The gas to solids ratio is constant in the whole disk (the embryos always stay beyond the ice line),
with $\Sigma_{\rm gas} / \Sigma_{\rm solids} = 70$.
The gas surface density evolves as a result of
viscous transport and photoevaporation:
$$
{d \Sigma \over d t} = {3 \over r} {\partial \over \partial r } \left[ r^{1/2} {\partial \over \partial r}
\nu \Sigma r^{1/2} \right] + \dot{\Sigma}_w(r)
.
$$
The photoevaporation term $\dot{\Sigma}_w(r)$ is taken as in Veras \& Armitage (2004).
The thermodynamical properties of the
disk as function of position and surface density (temperature, pressure, density scale height),
as well as the mean viscosity $\nu$, are calculated by solving the
vertical structure equations using the method presented in Papaloizou \& Terquem (1999) and  A05a. These
quantities are used to determine the composition of the ices incorporated in the planetesimals,
and finally the enrichments in volatile species in the two planets
(see Sect. \ref{volatiles}).

The key point in our models is
that both planets are formed concurrently within the same disk, hence the
physical assumptions and initial properties of the nebula are the same
for both. For a given disk model, we begin by searching a satisfactory
model matching the observed properties of Jupiter. Once such a model is
found, we try to adjust the two remaining parameters (initial location of
the embryo and time delay) to find a similarly suitable model for Saturn.
Our entire approach, as well as some tests we have  made to check our code,
can be found elsewhere (A05a), 
we give here some details on two points, the calculation of $M_{\rm
core}$ and $M_{\rm Z,enve}$, and the migration rates.

To estimate $M_{\rm core}$ and $M_{\rm Z,enve}$, we compute the fate of the
infalling planetesimals by computing their trajectory inside the envelope
as well as their mass loss. The latter results from thermal effects as
well as mechanical ablation due to Rayleigh-Taylor instabilities on the
planetesimals' surface (Korycansky et al. 2002).  This allows us to determine the fraction
of planetesimals' mass that directly reaches the core, which we identify as
$M_{\rm core}$, the core mass at the end of the formation process. The mass
deposited inside the envelope is assumed, due to convection\footnote{The interior of the planets is largely
unstable with regard to convection, even with the presence of molecular weight gradients: the radiative
gradient is dominant over the adiabatic one by orders of magnitude.
}, to
be homogeneously distributed within the envelope and is identified as
$M_{\rm Z,enve}$, the mass of heavy elements in the planet's envelope.
Finally, we note that processes like core erosion or settling could occur
during subsequent evolution of the planet and modify significantly the
values of $M_{\rm core}$ and $M_{\rm Z,enve}$ found here (see SG04). On the
other hand, the total mass of heavy elements in the planets should remain
constant.

The calculation of the planetesimal's trajectory also gives the place where
the energy of planetesimals is deposited. This quantity is used to calculate
the structure of the forming planet, by solving the standard internal
structure equations: the amount of energy released by infalling planetesimals
enters in the energy equation, in the sinking approximation (see P96 and A05a).

Low mass planets undergo type I migration at a rate being linear with the
planet's mass. However, the most recent analytical estimates of type I migration rates by Tanaka et al. (2002),
which have been derived assuming a laminar disk, are much too large
to be compatible with the observed frequency of extra-solar planets.
Therefore, planet survival implies a significantly reduced rate of type I
migration. First hints how this could be achieved have been obtained by Nelson \& Papaloizou (2004) in
numerical modelling of turbulent disks in which much reduced migration
rates have been obtained. In our calculations, we have reduced the rate
of type I migration by multiplying the analytical estimates by an arbitrary
factor $f_{\rm I}$, whose value varied in order to check its influence on the
results.
For higher mass planets, the migration is of type II (Ward 1997), the rate being independant
of the planet mass. When the mass of the planet becomes comparable to the one of the disk,
migration slows down and eventually stoppes. The switch from type I to type II occurs when the Hill's radius
of the planet is equal to the disk density scale height, which is calculated with the vertical structure
of the disk.
Finally, note that we do not take into account gravitational interactions between the two forming
planets that could alter the migration rates.

\section{Jupiter and Saturn formation}

We consider values of $f_{\rm I}$ between 0 (no type I migration) and 0.03
(as we shall see below, higher values would imply too large starting locations
of proto-Jupiter to account for the present structure of Saturn). For this range,
we find suitable Jupiters to form from embryos starting between 9.2 AU
(Astronomical Units) and 13.5 AU in a disk with a total mass ranging from
0.05 to 0.035 $M_\odot$ (solar masses)  and a total photoevaporation rate comprised
between 1 and 1.5 $\times 10^{-8} M_\odot /$ yr. For all the cases considered
here ($f_{\rm I}$=0, 0.001, 0.005, 0.01 and 0.03), it was possible to form within
3 Myr a planet whose final mass, location and global internal structure
were compatible with the Jupiter ones (see Fig. 1d). We note that the final
structure of the planet ($M_{\rm core}$, $M_{\rm Z,enve}$) is independant of
the assumed type I migration rate.  This rate only gives the starting location
of the embryo.

Concurrently with the formation of Jupiter, we also follow the growth of the proto-Saturn
embryo. The latter is started at a larger heliocentric distance and with an arbitrary
time delay. This implies that, depending upon initial conditions, proto-Saturn may
actually enter a region of the disk already visited and consequently modified (less
planetesimals for example) by proto-Jupiter (see Fig. 1c). In Fig. 2, we present
the successful Saturn formation model corresponding to the Jupiter model presented in
Fig. 1 (red curves, $f_{\rm I}$=0.001). The synthetic Saturn started as an embryo at
11.9 AU, 0.2 Myr after proto-Jupiter. The resulting planet exhibits characteristics
quite comparable to the actual Saturn (see Fig. 2d).
The mass of the core is slightly lower than the one allowed by
SG04. However, we recall that the mass derived in SG04 may be decreased by up 
to $\sim 7 \mearth$ due to the sedimentation of helium (see Sect. \ref{intro}).

The mass of Saturn's final core is similar to the one obtained for
Jupiter. This is because the core is built from the infalling
planetesimals that are able to traverse the gaseous envelope without
being disrupted. For a fixed envelope, disruption is essentially a
function of the size of the planetesimals which in our work is assumed
to be identical at all locations (100 km). Increasing the mass of the
planetesimals by a factor ten leads to core masses of the order of $8 \mearth$.

The importance of Jupiter's wake on the formation of Saturn is evidenced by the green
curves in Fig. 2a which were obtained by forming Saturn in the absence of Jupiter in
an otherwise identical disk. In this case, the increased rate of planetesimal infall
prevented the accretion of a sizeable envelope, and the resulting planet, while at
Saturn's current location, remained quite small ($20 M_\oplus$). This can be explained by
the fact that the gas accretion rate is inversely dependent upon the energy deposited
by infalling planetesimals. Thus, once proto-Saturn's feeding zone enters a region
previously depleted in planetesimals by the passage of Jupiter, their infall is
reduced and gas accretion proceeds at a faster rate, ultimately leading to a more
massive planet than in the case without planetesimal depletion. In the latter case
(ignoring the effects of Jupiter's formation), we checked that even by varying the
initial location and formation starting time, it was not possible to obtain a
Saturn-mass planet at its current location (see Fig. 2c).
Increasing the starting location
of Jupiter (beyond $\sim$ 10 AU, corresponding to $f_{\rm I}$ larger than 0.01),
results in Saturn-like planets containing
too few heavy elements compared to the actual planet.
Moreover, the mass of accreted planetesimals never reaches a
level which could trigger a significant accretion of gas. At the present
location of Saturn, the synthetic planet remains less massive than the
actual one (see Fig 2c, blue curve).

\section{Enrichments in volatile species}
\label{volatiles}

We now concentrate on the models that can form both Jupiter and Saturn
(red ones in Fig. 1a and Fig. 2a) and examine whether they can also account for
the volatile abundances measured in the atmospheres of the two planets.
To do this, we use the clathrate trapping theory (Lunine \& Stevenson 1985) and the
thermodynamical conditions inside the disk as calculated in our models to
compute the composition of ices incorporated in the planetesimals. Knowing
the mass of accreted planetesimals, we compute the total expected abundances
of some volatile species, in our case C, N, S, Ar, Kr, and Xe.

We use for this calculation the recent solar abundances determinations of Lodders (2003).
C is set to have been present in the solar nebula vapor phase under the form
of CO$_2$, CO and CH$_4$, with CO$_2$:CO:CH$_4$=30:10:1, ratios which are
compatible with ISM measurements (see Allamandola et al. 1999, Gibb et al. 2004). Moreover, N is taken to have
been present under the form of N$_2$ and NH$_3$, with a ratio NH$_3$:N$_2$=1,
and S under the form of H$_2$S and other sulfur compounds (Pasek et al. 2005). Other
initial ratios of CO$_2$:CO:CH$_4$ and NH$_3$:N$_2$ can lead to slightly
different abundances of volatiles, but do not modify our main conclusions.
We note finally that CO$_2$ crystallizes as a pure condensate prior to be
clathrated (see Alibert et al. 2005b) which has a considerable influence on the total
amount of water required for trapping all the volatiles in the planet.

The results of these abundances calculations in the case of Jupiter have been
presented elsewhere in details (see Alibert et al. 2005b), we only summarize the main conclusions
here: C, N, S, Ar, Kr and Xe are enriched respectively by a factor of about 2.8, 2.5,
2.1, 2, 2.1, 2.6 compared to their solar values. These values are compatible
with the {\it in situ} measurements made by the {\it Galileo} probe and recalled at the beginning of this Letter.
The resulting enrichment for oxygen (not yet measured) is at least O/H =
$3.4 \times 10^{-3}$ or $\sim 6$ times the solar value.

In the case of the Saturn formation model, we obtain enrichments of 2.4 and
2.2 compared to solar values for C and N respectively. This is again
compatible with the ground-based observations quoted in the introduction. Moreover, we predict
that S, Ar, Kr, and Xe are enhanced by a factor of respectively 1.9, 1.7,
1.9 and 2.3.  Our formation model predicts the accretion of $\sim 13.2
M_\oplus$ of heavy elements, and the trapping of the volatiles result from
the accretion of at least $5.4 M_\oplus$ of ices (depending on the efficacity
of the clathration process).  These two calculations imply that the mean
Ices/Rocks (I/R) ratio of accreted planetesimals was $> 0.7$, a value
consistent with the one inferred for Saturnian satellites.  Finally, the
resulting enrichment of O in Saturn is O/H $\sim 3 \times 10^{-3}$, ie 5.2
times the solar value.

\section{Summary and discussion}

We have calculated in this Letter formation models of the two gas giant planets
of our Solar System, in the framework of our extended core-accretion models
taking into account migration and disk evolution. 
The calculations presented here are simplified in some aspects, that could
be improved in the future. In particular,  our disk model is calculated in the
framework of the $\alpha$ formalism of Shakura \& Sunyaev (1973), which itself
is a limitation. Moreover, we do not take into account
gravitational interactions between the two planets that can alter the migration
rates.

Our calculations allowed us to give an estimate of
the core mass, enrichment in heavy elements, and
enrichment in volatile species that can be compared with observational
data about Jupiter and Saturn. 
These calculations therefore show that our models
can lead to the formation
of two giant planets closely resembling our Jupiter and Saturn in less than
3 Myr. In order for our synthetic planets to match the bulk properties of the
two gas giants in our solar system, we found that Jupiter must have started
at a heliocentric distance smaller than $\sim$ 10 AU otherwise Saturn, which
follows in its trail, cannot accrete enough heavy elements.  The heavy elements
content as well as the core mass obtained for our synthetic planets are in
good agreement with the interior models of SG04.

Finally, in the framework of our model, the enrichments
(compared to solar) of some volatile species measured in the atmosphere of
both giant planets can also be accounted for
in a self-consistent manner.
However, we note that, recently, the {\it Cassini} spacecraft measurements have led to a revised
value of the abundance of C in Saturn's atmosphere of $8.1 \pm 1.6$ times
the solar value (see Flasar et al. 2005).  Using the clathrate hydrate trapping theory,
and assuming the minimum value for the C abundance (6.5 times the solar
value), we obtain abundances of N, S, Ar, Kr, and Xe of  respectively 5.9,
5, 4.7, 5.1 and 6.1 times the solar values. The resulting O enrichment would
be about 14 times the solar value.  These predicted enrichments are
significantly higher than the ones quoted in Sect. \ref{intro},
in particular for N.  The future confirmation of both the new measurement of
C and of the "old" value of N would imply that there has been some unknown
fractionation processes between these species in the solar nebula gas-phase,
and that their abundances in Saturn cannot be explained using solely the
standard clathrate hydrate trapping theory. However, the measurement of C
in Saturn by the {\it Cassini} spacecraft may be subject to revisions in a near
future.

\acknowledgements 
This work was supported in part by the Swiss National Science Foundation.  O.M. was 
supported by an ESA external fellowship.

{}

\clearpage

\begin{figure}
\begin{center}
\centerline{\includegraphics[width=10cm]{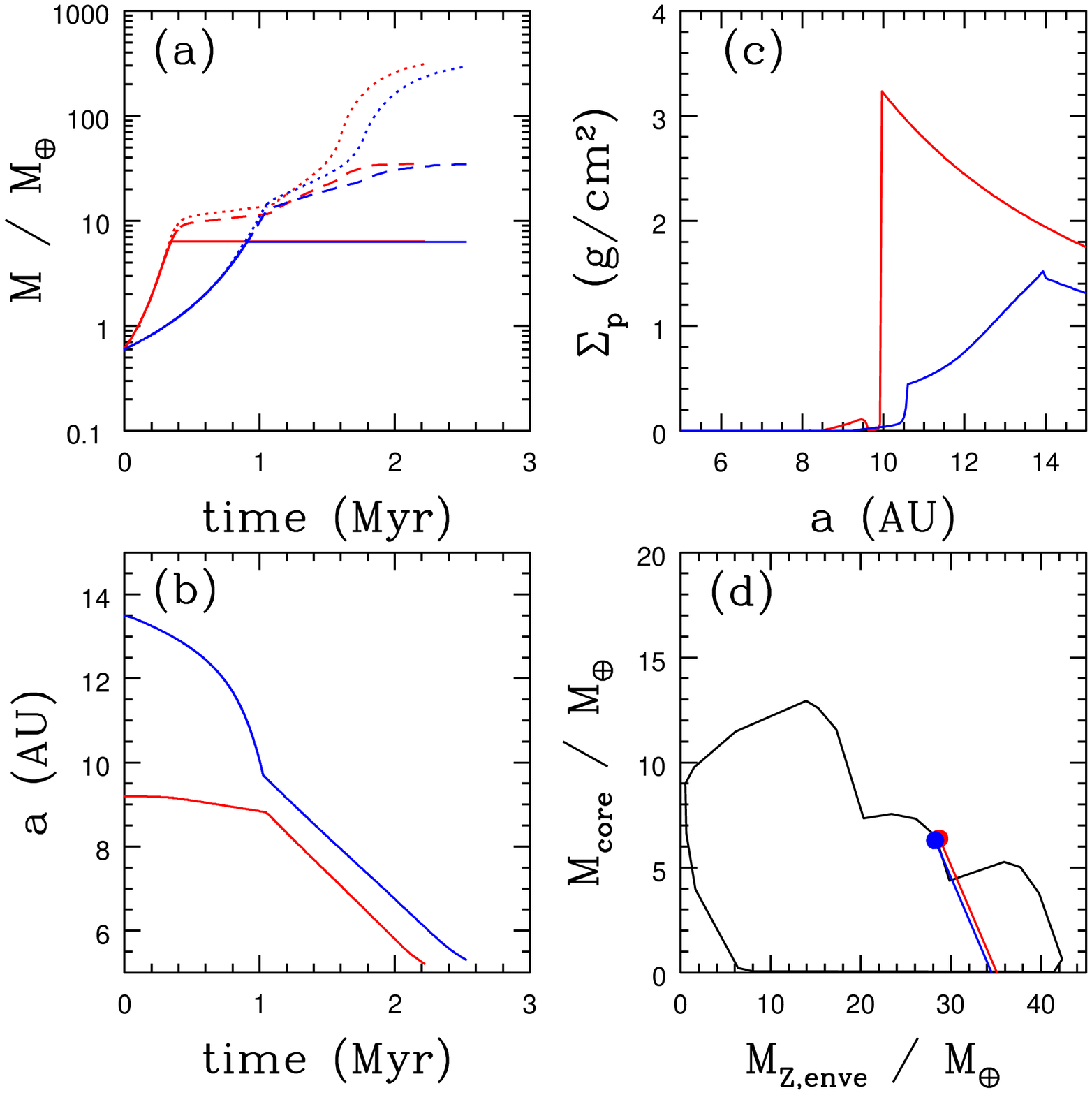}}
\caption{
Jupiter formation models. The red (resp. blue) curves are obtained using $f_{\rm I}$ = 0.001
(resp. 0.03). (a) Mass of accreted planetesimals (dashed lines), total mass (dotted lines)
and mass of solid core (solid lines) as a function of time for two simulations.
(b) Heliocentric distance as a function of time. (c) Final surface density of planetesimals 
$\Sigma_p$ as a function of heliocentric distance. (d) Core mass ($M_{\rm core}$) and mass of heavy elements
($M_{\rm Z,enve}$) dissolved in the envelope. The black curve gives the domain allowed by the present
day structure models of SG04. If some core dissolution occurs after the formation, the two points
would evolve along the lines.
For all the values of $f_{\rm I}$ we used, the points representing the final structures are
very close to the red and blue ones. These points are not represented here for clarity.
}
\end{center}
\label{J}
\end{figure}

\clearpage

\begin{figure}
\begin{center}
\centerline{\includegraphics[width=10cm]{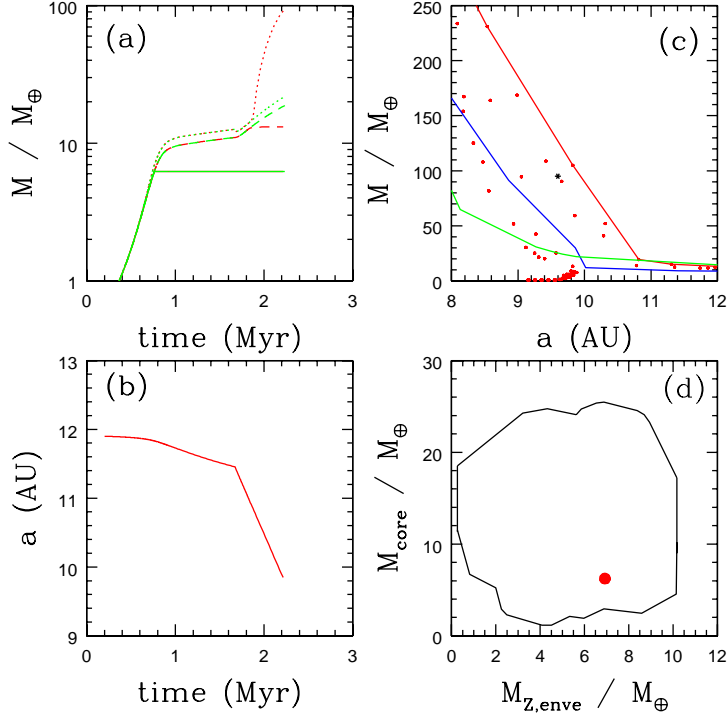}}
\end{center}
\caption{
Saturn formation models. The red (resp. green) lines are obtained while taking
into account the effect of Jupiter (resp. without this effect). (a,b)  Same as
Fig. 1 a,b. (c) Red small dots: different Saturn models (final mass and position), varying
the initial position and the time delay after the beginning of Jupiter's formation.
The approximate maximum mass that can be reached by proto-Saturn at a given distance
to the sun is indicated by the lines.
The blue line is similar to the red one, except that the
corresponding Jupiter formation is calculated with $f_{\rm I} = 0.03$ (blue curves in Fig. 1).
The green line is similar to the red one, except that the effect of Jupiter's wake on Saturn formation
is {\it not} taken into account.
The black star indicates Saturn's position
in this diagram.
(d)
Core mass ($M_{\rm core}$) and mass of heavy elements
($M_{\rm Z,enve}$) dissolved in the envelope. The black curve gives the domain allowed by the present
day structure models of SG04, taking into account a possible shell of sedimented helium,
of mass between $0$ and $7 M_\oplus$.
}
\label{S}
\end{figure}

\end{document}